\documentclass[
  a4paper,
  twocolumn, 
  pra,
  superscriptaddress,
  amsfonts,amsmath,amssymb,
  longbibliography
]{revtex4-2}

\usepackage[
  colorlinks,
  citecolor=blue,
  linkcolor=blue,
  urlcolor=blue
]{hyperref}
\usepackage{graphicx,epstopdf,bm}

\usepackage{braket}
\usepackage[capitalize]{cleveref}
\usepackage{xcolor}
\usepackage[all]{hypcap}
\usepackage{multirow}

\crefname{section}{Sec.}{Secs.}
\Crefname{section}{Section}{Sections}
\crefname{appendix}{App.}{Appces.}
\Crefname{appendix}{Appendix}{Appendices}

\DeclareMathOperator{\Tr}{Tr}
\DeclareMathOperator{\Real}{Re}
\DeclareMathOperator{\Imag}{Im}
\newcommand{\e}[1]{\mathrm{e}^{#1}}
\newcommand{\op}[1]{\hat{#1}}

\newcommand{\cf}{cf.\ }
\newcommand{\eg}{e.g.\ }
\newcommand{\ie}{i.e.\ }

\newcommand{\figref}[2][]{Fig.~\hyperref[#2]{\ref*{#2}#1}}
\newcommand{\figureref}[2][]{Figure~\hyperref[#2]{\ref*{#2}#1}}

\newcommand{\sqrtiswap}{$\sqrt{i\mathrm{SWAP}}$}

\newcommand{\todothesis}[1]{}

\begin{document}

\title{The perfect entangler spectrum as a tool to analyze crosstalk}


\author{Matthias G. Krauss}
\author{Christiane P. Koch}
\affiliation{Freie Universit\"{a}t Berlin, Fachbereich Physik and Dahlem Center for Complex Quantum Systems, Arnimallee 14, 14195 Berlin, Germany
}


\begin{abstract}
    Crosstalk is a key obstacle to scaling up quantum computers. It may arise from persistent qubit-qubit couplings or dynamically during gate operation, with the latter being particularly difficult to detect. Here, we introduce the perfect entangler spectrum as a means to identify dynamic crosstalk leading to undesired entanglement. It leverages the geometric classification of two-qubit gates in terms of perfect entanglers. We exemplify application of the spectroscopy for fixed-frequency transmons and parametrically driven gates: When scanning the frequency of a spectator qubit, peaks in the perfect entangler spectrum signal dynamic crosstalk, and analysis of the peaks reveals the mechanisms causing the crosstalk. We discuss the experimental implementation of the crosstalk spectroscopy which requires two two-qubit gate tomographies.
\end{abstract}


\maketitle


\section{Introduction}
  \label{sec:introduction}

Quantum computing relies on well-characterized, individually addressable qubits that can be initialized, read out and operated in the desired way~\cite{DiVincenzoFP00}.
A major challenge in many platforms is crosstalk which describes the unwanted interactions of different parts within a quantum processor~\cite{SarovarQ20}.
It can be divided into two categories, persistent crosstalk due to constant interactions between different parts of the quantum processor, and crosstalk that occurs during operations. The latter can be caused directly by the presence of other qubits or by controls acting on more than the intended qubits. Hardware architectures that rely on frequency to address the qubits are particularly prone to crosstalk errors, and  the problem becomes more severe as the number of qubits is scaled up due to frequency crowding. Examples are superconducting qubit architectures~\cite{AshSakiITQE20,KettererPRA23} and trapped ions~\cite{Strohm24}. In superconducting qubits, control is exerted via macroscopic electromagnetic fields that easily leak over to adjacent qubits~\cite{AbramsPRA19,KosenPQ24}. 
In ion chains, crosstalk is a problem even in architectures with single-site addressability, 
since entanglement generation relies on the shared normal modes of vibrations in the trap which get more dense for a larger number of ions~\cite{LandsmanPRA19,SchwerdtPRX24}. 
  
In recent years, significant efforts have been taken to detect~\cite{SarovarQ20,RudingerPQ21}, characterize~\cite{GambettaPRL12,RudingerPQ21,KettererPRA23,Zhou25},  and mitigate~\cite{MundadaPRA19,ParradoRodriguezQ21,SungPRX21,KandalaPRL21,FlanneryQST25} crosstalk.
Detection of crosstalk often uses a variant of randomized benchmarking~\cite{GambettaPRL12,McKayPRL19,SarovarQ20,KettererPRA23}, whereas mitigation approaches include optimization of both the architecture design~\cite{MorvanPRR22,Zhang24,Ai25} and the pulses which implement gates~\cite{ZhouPRL23,WatanabePRA24,YouPRA24}.
While the focus has been on coherent crosstalk errors, correlated noise spectroscopy can be used to assess the impact of  an environment~\cite{VonLupkePQ20,Greggio25}.
For superconducting quantum computing, mainly ZZ-crosstalk has so far been analyzed~\cite{MundadaPRA19,KandalaPRL21,StehlikPRL21,ZhangPQ24,Lange25}. It originates from unwanted, persistent couplings which result in undesired entanglement even if no gate is applied.
By improving hardware designs and utilizing new types of superconducting couplers, ZZ-crosstalk has been mitigated effectively~\cite{MoskalenkonQI22,HeunischPRA23,ZhangPQ24,Ai25}.
In contrast to such persistent crosstalk, crosstalk during gate operation, due to the undesired interaction with other qubits or due to signal leakage, is more difficult to detect and mitigate~\cite{DaiPQ21,Balewski25}.
Of particular interest is the influence of spectator qubits on the gate operations. For example for superconducting architectures, 
the importance of understanding and mitigating the impact of idle spectator qubits in has been highlighted~\cite{KrinnerPRA20,MalekakhlaghPRA20,CaiPRL21,ZhaoPQ22}.
However, most analysis methods are tailored to a single two-qubit gate, such as the CR gate~\cite{KrinnerPRA20,MalekakhlaghPRA20,CaiPRL21} or do not consider excited spectator qubits~\cite{Ai25}.
Furthermore, many protocols consider any deviation from the gate protocol as crosstalk, but different deviations are more or less easy to correct.
It is thus useful to distinguish two types of deviations. The first type affects the entangling power of the two-qubit gate, i.e., it leads to gates that are no longer perfectly entangling. The second type comprises deviations leading only to single qubit rotations, which leave the gate perfectly entangling. This distinction is crucial since the entangling power is the primary resource of the two-qubit gate and any deterioration leads to the necessity of more two-qubit gates. On the other hand, single qubit rotations can be corrected effectively or accounted for in a compiling step. Ref.~\cite{MalekakhlaghPRA20} addresses this issue and analyzes the entangling power for the CR gate between two qubits, but surprisingly this step is missing for their consideration of additional spectator qubits.

Here we introduce the ``perfect entangler spectrum'' to detect crosstalk in qubit gate operations that arises from the presence of additional qubits. 
We focus on crosstalk that results in an imperfect entangling gate for two qubits due to the presence of other qubits that are not expected to participate in the gate operation. Two-qubit entangling gates are a fundamental building block of quantum computation, in particular in the context of quantum circuits and universal quantum computation~\cite{Nielsen10}. They are particularly prone to crosstalk since engineering an interaction between two qubits can easily result in unwanted interactions with other, nearby qubits. Such crosstalk can manifest in different forms:  Either the gate is no longer capable to perfectly entangle the targeted qubits or the gate generates entanglement with non-targeted qubits. Focussing on this type of crosstalk, we employ the geometric classification of two-qubit gates in terms of local invariants~\cite{MakhlinQIP02,ZhangPRA03} to derive a protocol for its detection. To this end, we extend a measure that quantifies how far a two-qubit gate is from the set of all perfect entanglers~\cite{WattsPRA15,GoerzPRA15} to operations involving also a third -- spectator -- qubit. The value that the extended measure takes as a function of the spectator frequency is what we call the  perfect entangler (PE) spectrum.

To illustrate the use of our protocol, we consider fixed-frequency transmons that interact via a tunable coupler~\cite{McKayPRA16,ReagorSA18,YanPRA18,AruteN19,StehlikPRL21,SungPRX21,DingPRX23,ZhangPQ24,Zhang25}.
For two popular, parametrically driven gates, the \sqrtiswap{} and the controlled phasegate~\cite{MajerN07,McKayPRA16,ReagorSA18,AruteN19,SetePRA21,SungPRX21,DingPRX23,GoogleQuantumAIN23,ZhangPQ24,Huber24},  
we show that, when crosstalk occurs, the PE spectrum displays peaks as a function of the frequency of a spectator qubit. Analysis of the peaks allows us to identify the mechanism that leads to the crosstalk. Importantly, we find that crosstalk occurs not just at obvious resonances that can be read off the Hamiltonian but also through complex interactions with the pulse that lead to a variety of crosstalk pathways. 
Finally, we discuss how PE spectroscopy can be realized in experiment.


\section{Theoretical Framework}
  \label{sec:theoretical-framework}

 \subsection{The Perfect Entangler Spectrum}
    \label{ssec:the-perfect-entangler-spectrum}

Our starting point is a measure that quantifies, for a given gate acting on two qubits, its distance to the ``space'' of perfect entanglers. These are all gates capable of generating maximally entangled states from initially separable states~\cite{ZhangPRA03}. The measure has been constructed as an optimization functional in quantum optimal control theory to target an arbitrary entangling two-qubit operation rather than a specific gate~\cite{WattsPRA15,GoerzPRA15},
  \begin{align}
    \tilde{J}_{\mathrm{PE}} =
      \Big(g_3\sqrt{g_1^2+g_2^2}-g_1\Big)\,.
    \label{eq:perfect-entangler-functional-plain}
  \end{align}
Here, $g_1$, $g_2$, and $g_3$ are the so-called local invariants~\cite{MakhlinQIP02,ZhangPRA03}, the theory of which is recalled for completeness in App.~\ref{app:LI}.
$\tilde{J}_{\mathrm{PE}}$ goes to zero as the space of PEs, the so-called polyhedron of PEs, cf.\ App.~\ref{app:LI}, is approached. Note, that this is true for both sides of the polyhedron boundary such that it is necessary to ascertain whether initially the gate coordinates are already within the polyhedron, in which case it is a perfect entangler. If the gate is located outside the polyhedron, $\tilde{J}_{\mathrm{PE}}$ provides a measure for the proximity between the gate and the closest perfect entangler.

The framework of local equivalence classes and thus \cref{eq:perfect-entangler-functional-plain} is only valid for unitary evolutions within $\mathrm{SU}(4)$. Typically, however, the logical subspace is embedded in a larger Hilbert space, and the evolution within the logical subspace may be non-unitary. To quantify the non-unitarity, $\tilde{J}_{\mathrm{PE}}$ is amended by an additional term~\cite{GoerzPRA15}:
  \begin{subequations}\label{eq:PEfunc}
  \begin{align}
    J_{\mathrm{PE}} =
      (1-w_U)\big(g_3\sqrt{g_1^2+g_2^2}-g_1\big)
      + w_U \Delta_U\,,
    \label{eq:perfect-entangler-functional}
  \end{align}
  with
  \begin{align}
    \Delta_{U}=\Big(1-\Tr\left[\op{U}^\dagger\op{U}\right]/4\Big)\,.
    \label{eq:delta_u}
  \end{align}
  \end{subequations}
$\Delta_U$ leads to an increase of the functional value in case the evolution is non-unitary. The weight $w_U$ should be chosen large, since the first term becomes meaningless for strongly non-unitary evolutions~\cite{GoerzPRA15}.

We now generalize \cref{eq:PEfunc} to three qubits for which all ideal gates are elements of $\mathrm{SU}(8)$. If there is no crosstalk, the third qubit is not influenced by the entangling operation between the first two qubits. Thus, the logical subspace can be decomposed into two $\mathrm{SU}(4)$ subspaces, one for each basis state of the third qubit. If the two-qubit gate is executed without any influence of the third qubit, the two-qubit gates in the two subspaces must be the same. 
To verify the similarity of the gates in the logical subspaces for the third qubit being in $\ket{0}$ and $\ket{1}$, we introduce
  \begin{align}
    \mathcal{S} = 1- \bigg|
        \!\Tr\Big[
          \big(U^{(q_3=\ket{0})}(T)\big)^\dagger U^{(q_3=\ket{1})}(T)
        \Big]/4
      \bigg|^2 \,,
    \label{eq:similarity-functional}
  \end{align}
where $U^{(q_3=\ket{0})}(T)$ and $U^{(q_3=\ket{1})}(T)$ are the two-qubit gates in the two subspaces. Combining 
$\mathcal{S}$ with the PE functional evaluated in each of the two subspaces, we obtain a measure for how well two qubits are entangled without affecting a third qubit,
  \begin{align}
    J = \Big(
        J^{(q_3=\ket{0})}_{\mathrm{PE}} + J^{(q_3=\ket{1})}_{\mathrm{PE}}
        + w_S\mathcal{S}
      \Big)\,.
    \label{eq:full-functional}
  \end{align}
Here, $w_S$ is a weight for the similarity measure, relative to the first two. Note that $\mathcal{S}$ is phase-agnostic: Any relative phase between the two gates is not detected. Such a relative phase corresponds to a local $Z$-rotation of the third qubit that does not impair the entanglement of the first two qubits. In other words, $J$ quantifies whether the two qubits on which the entangling operation (no matter which one precisely) is executed remain unaffected by a third qubit. To assess when the presence of the third qubit causes crosstalk, we vary its frequency and determine for each $\omega_3$ the smallest value of $J$ throughout the protocol. The resulting curve is the ``PE spectrum''.  

  \subsection{Tunable Coupler Architecture}
    \label{ssec:tunable-coupler-architecture}

  To exemplify the application of the PE spectrum for identifying crosstalk, we consider
  a system of three fixed-frequency transmons, which are coupled via a central transmon with tunable frequency, acting as coupler~\cite{McKayPRA16},
  \begin{align}
    \op{H}&=\omega_c(t) \op{b}^\dagger\op{b}
        - \frac{\alpha_c}{2} \op{b}^\dagger\op{b}^\dagger\op{b}\op{b}
        +\sum^3_{j=1} \Big[\omega_j \op{a}_j^\dagger\op{a}_j
        - \frac{\alpha_j}{2}\op{a}_j^\dagger\op{a}_j^\dagger\op{a}_j\op{a}_j
      \Big]
    \notag\\
      &\qquad
      +\underbrace{ \sum^3_{j=1}g_j \big(\op{b}+\op{b}^\dagger)\big(\op{a}_j
      + \op{a}_j^\dagger\big)}_{\op{H}_{Q-C}} \,.
    \label{eq:tunable-coupler-hamiltonian}
  \end{align}
  Here, $\op{b}$ is the annihilation operator of the tunable coupler and $\op{a}_j$ are the annihilation operators of the fixed-frequency transmons.
  The logical states are the eigenstates of the uncoupled transmon qubits, with the tunable coupler being in the ground state,
  $\ket{n_1, n_2, n_3, n_{\mathrm{tc}}=0} \equiv \ket{n_1n_2n_3}$.

Modulating the frequency of the central coupler allows for driving  transitions between the qubits which leads to controllable interactions~\cite{BertetPRB06,NiskanenS07}. This feature forms the basis for implementing various gates between the qubits. For flux tunable couplers, the frequency typically varies with the flux $\Phi$
  as~\cite{KochPRA07}
  \begin{subequations}
    \begin{align}
      \label{eq:time-dependent-freq}
      \omega_c(t) &= \omega^{\mathrm{max}}_c
      \sqrt{\big|\cos{\big(\pi\,\Phi(t)\big)}\big|}\,,
    \end{align}
    where
    \begin{align}
      \Phi(t) &= \Theta + \delta\cos(\omega_\phi t)\,.
      \label{eq:time-dependent-flux}
    \end{align}
    \label{eq:time-dependent-pulse}
  \end{subequations}
Such single-frequency flux drives $\Phi(t)$ have been used in parametrically driven two-qubit gates~\cite{McKayPRA16,CaldwellPRA18,ReagorSA18,GanzhornPRR20,PetrescuPRA23}. They result in  (approximately) two non-zero frequency components $\omega_\phi$ and $2\omega_\phi$ in $\omega_c(t)$. To model the experimental turn on and off process, we use a Gaussian flattop  as envelope with width $\sigma_t$.

Below we study exemplarily a controlled phasegate (also referred to as controlled-Z (CZ)) and a \sqrtiswap{} gate.
In preparation of our analysis of the PE spectrum, it is instructive to recall the mechanisms that implement the gates.
The CZ gate is obtained via a population swap between $\ket{11}$ and $\ket{20}$, with the population in all other logical states staying constant~\cite{ReagorSA18}. Upon the transition to $\ket{20}$ and back, the state $\ket{11}$ acquires a phase $\phi$, relative to all other logical states. Tuning this phase to be $\pi$ yields CZ.  The second gate, \sqrtiswap{}, is implemented by driving the tunable coupler close to the resonance frequency of the $\ket{01}\leftrightarrow\ket{10}$ transition~\cite{MajerN07,McKayPRA16}. This induces a partial swap together with a relative phase between these states. 

For both protocols, the tunable coupler stays in its ground state during  gate execution; 
its role is merely to mediate the coupling between the qubits which, by design, are not directly coupled. Although
the drive $\omega_c(t)$ only modulates the coupler frequency, 
in the dressed basis, with the dressing due to the qubit-coupler interaction $\op{H}_{Q-C}$, it obtains an off-diagonal component ("$\op{X}$") and thus can induce transitions. We refer to those as first order $X$-transitions.
Transitions between logical states occur at second and higher order perturbation theory with respect to the drive $\omega_c(t)$. In second order perturbation theory, transitions may involve both diagonal and off-diagonal components of the drive ($XZ$ or $ZX$) or correspond to two simultaneous $X$-transitions. We refer to the latter as second order $X$-transitions.
A detailed discussion of which transitions occur at what perturbative order are provided in \cref{app:diagonal-drive-transitions,app:perturbation-theory-second-order}.


\section{Analyzing Crosstalk Mechanisms: Application to a Controlled-Z Gate}
  \label{sec:analysing-crosstalk-mechanisms-cz}

\begin{table}[tbp]
    \caption{Parameters for CZ gate~\cite{GanzhornPRR20}.}
    \label{tab:parameters-table-ganzhorn}
    \begin{ruledtabular}
      \begin{tabular}{llcr}
        \multirow{3}{4em}{Qubit 1}
          & Frequency & $\omega_1/2\pi$ & $5.089\,\mathrm{GHz}$\\
          & Anharmonicity & $\alpha_1/2\pi$ & $310\,\mathrm{MHz}$\\
          & Coupling & $g_{1}/2\pi$ & $116\,\mathrm{MHz}$\\
      \colrule
        \multirow{3}{4em}{Qubit 2}
          & Frequency & $\omega_2/2\pi$ & $6.189\,\mathrm{GHz}$\\
          & Anharmonicity & $\alpha_2/2\pi$ & $286\,\mathrm{MHz}$\\
          & Coupling & $g_{2}/2\pi$ & $142\,\mathrm{MHz}$\\
      \colrule
        \multirow{3}{4em}{Qubit 3}
          & Frequency & $\omega_3/2\pi$ & varied\\
          & Anharmonicity & $\alpha_3/2\pi$ & varied\\
          & Coupling  & $g_{3}/2\pi$ & $85\,\mathrm{MHz}$\\
      \colrule
        \multirow{2}{4em}{Tunable coupler}
          & Frequency & $\omega^{\mathrm{max}}_c/2\pi$ & $8.1\,\mathrm{GHz}$\\
          & Anharmonicity & $\alpha_c/2\pi$ & $235\,\mathrm{MHz}$\\
      \colrule
        \multirow{4}{4em}{Drive}
          & Offset & $\Theta$ & $0.15$\\
          & Amplitude & $\delta$ & $0.19$\\
          & Frequency & $\omega_\phi/2\pi$ & $816.58\,\mathrm{MHz}$\\
          & On/Off flank width & $\sigma_t$ & $13\,\mathrm{ns}$\\
      \end{tabular}
    \end{ruledtabular}
  \end{table}
  
 \begin{figure*}[tbp]
    \includegraphics[width=\textwidth]{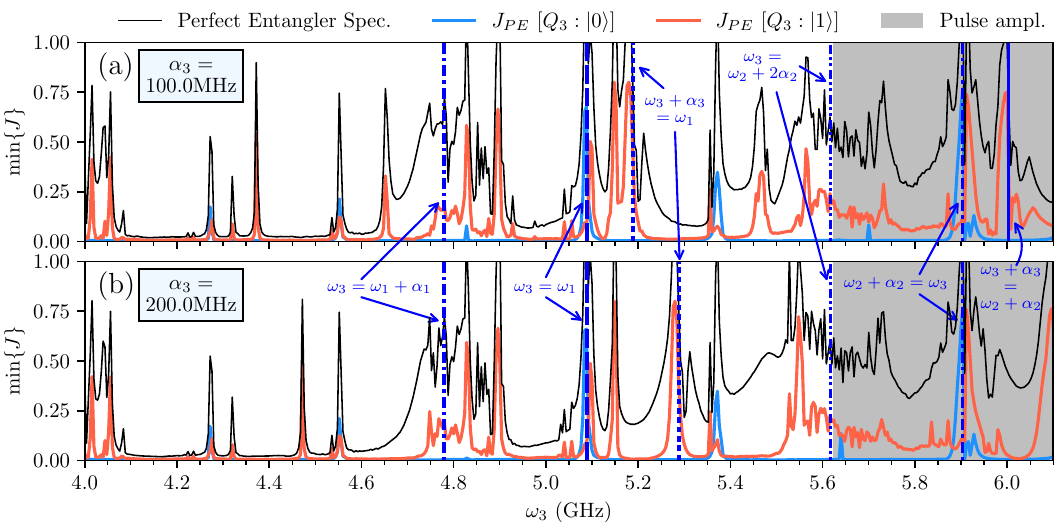}
    \caption{
      PE spectrum, i.e., value of $J$, \cref{eq:full-functional}, obtained when
      varying the spectator frequency for two 
      spectator anharmonicities $\alpha_3$ (black line) and the spectra conditional on the spectator state being 
      $\ket{0}$ (blue), resp. and $\ket{1}$ (orange). 
      Note that the black curve is not simply the sum of the blue and orange
      curves, but also contains the measure for the similarity of the
      two-qubit gates in the two subspaces (\cf \cref{eq:similarity-functional}).
      The vertical lines indicate resonance conditions.
    }
    \label{fig:pe_spectrum}
  \end{figure*}
  
We now show  how the PE spectrum signals crosstalk, starting with the example of the CZ gate and the parameters of Ref.~\cite{GanzhornPRR20}, cf. \cref{tab:parameters-table-ganzhorn}. As suitable weights in \cref{eq:perfect-entangler-functional}, resp.  \cref{eq:full-functional}, we heuristically find $w_U=0.8$ and $w_S=0.5$ which allow us to readily identify all relevant features in the PE spectrum. For each spectator frequency $\omega_3$, we solve the time-dependent Schr\"odinger equation for all logical basis states as initial state and calculate the value of the functional $J$~\eqref{eq:full-functional}.

\figureref{fig:pe_spectrum} displays the minimum of $J$ over time  plotted against $\omega_3$ (black solid line) for two different anharmonicities $\alpha_3$ of the spectator qubit.  Peaks in the resulting spectrum signal a failure of the protocol due to crosstalk.  Small values, on the other hand, indicate the implementation of a perfectly entangling gate between qubits 1 and 2, without being affected by the spectator qubit. In other words, regions with small values of $J$ can be considered crosstalk-free.
The gray area in \cref{fig:pe_spectrum} indicates the range in which the frequency of the tunable coupler $\omega_c(t)$ is varied. The increase in overall amplitude of the spectrum within and close to this region reflects the dressing that  is caused by the drive and that leads to  a distortion of the protocol.

In addition to the full perfect entangler spectrum, the first and second term  in \cref{eq:full-functional} are shown as blue and red lines, respectively. Considering these contributions separately allows for identifying the type of crosstalk. For example, there are regions where the PE spectrum displays large peaks whereas the red and blue curves are almost zero, \eg
around $4.7\,\mathrm{GHz}$ in both panels of \cref{fig:pe_spectrum} or around $5.45\,\mathrm{GHz}$ for  $\alpha_3=200\,\mathrm{MHz}$ in \cref{fig:pe_spectrum}(b). The small values of the red and blue curves indicate the correct implementation of a PE in each of the subspaces with the spectator qubit in $\ket{0}$ and $\ket{1}$, respectively. The large value of the PE spectrum is then due to the last term $\mathcal{S}$.  In this case, the two-qubit gates in the two subspaces are not identical, which signals unwanted entanglement with the spectator qubit.

We will see below the occurrence of resonances between energy levels is the main reason for crosstalk. There are two types of resonances --  static and drive-induced ones. We use the term static resonance to refer to two or more transitions having the same transition frequency. These resonances result in population exchange, even without external drive. They are easily inferred from the Hamiltonian of the uncoupled transmons, without any knowledge of the dynamics. By contrast, drive-induced resonances are transitions that occur only when the drive is on. They are mediated by coupling between the transmons and can be understood with low order perturbation theory. In the following, we analyze both types of resonances in detail.

\subsection{Static Resonances}
    \label{ssec:static-resonances}

Typically, the qubit parameters are chosen to avoid static resonances, but adding more qubits introduces additional levels which may lead to static resonances. In \cref{fig:pe_spectrum}, we indicate the static resonances by vertical blue lines. The strongest static resonances appear if the spectator frequency is the same as one of the other qubits', i.e., $\omega_3 = \omega_{1/2}$  (dashed vertical lines in \cref{fig:pe_spectrum}). These resonances do not necessitate a dedicated spectroscopy as they are readily understood as addressability problem associated with the degeneracy of qubit frequencies.

Another type of static resonance involves higher levels of the transmons, beyond the logical subspace. Resonances due to transitions between $\ket{1}\leftrightarrow\ket{2}$ (dotted and dash-dotted vertical lines in  \cref{fig:pe_spectrum}) clearly coincide with high peaks in the spectrum indicating strong crosstalk.
Transitions involving the second excited states of the spectator (dotted vertical lines in \cref{fig:pe_spectrum}), vary between the two panels, as expected for different anharmonicities $\alpha_3$. Note that it is not sufficient to consider resonances involving transitions $\ket{0}\leftrightarrow\ket{1}$ in one qubit and $\ket{1}\leftrightarrow\ket{2}$ in another qubit. For example, in the CZ gate protocol, population may cycle through various states $\ket{02k}$ where $k$ labels the spectator state. This enables transitions involving higher energy levels, such as $\ket{021}\leftrightarrow\ket{030}$ or $\ket{021}\leftrightarrow\ket{012}$. In \cref{fig:pe_spectrum}, these transitions are indicated by dash-dotted lines near  $5.6\,\mathrm{GHz}$ and by a solid line near $6.0\,\mathrm{GHz}$ for  $\alpha_3=100\,\mathrm{MHz}$ (for $\alpha_3=200\,\mathrm{MHz}$, the latter is shifted out of the plotted frequency range). The transitions near $6.0\,\mathrm{GHz}$ align with a peak in the spectrum which shifts with $\alpha_3$. In contrast, for the transition around $5.6\,\mathrm{GHz}$, we do not observe a single well-defined peak. Instead, a larger region is affected by substantial crosstalk. The corresponding feature in the spectrum cannot be understood by considering only static resonances, but become clear when considering drive-induced resonances below.

  \subsection{Drive-Induced Resonances}
    \label{ssec:drive-modulated-resonances}

In order to quantify the drive-induced resonances, 
we first need to know which transitions can possibly occur. Analysis of a simpler model, cf. \cref{app:diagonal-drive-transitions}, suggests that the flux drive from \cref{eq:tunable-coupler-hamiltonian} induces transitions between levels which are coupled by $\op{H}_{Q-C}$. Furthermore, the effective drive is comparatively strong, which enables higher order $X$-transitions as detailed in \cref{app:perturbation-theory-second-order}.
This is necessary, since the gate mechanisms discussed in \cref{ssec:tunable-coupler-architecture} rely on second order $X$-transitions.

Based on these observations, we define a resonance measure that collects the information on all $X$-transitions of a given order $n$, that are allowed, in terms of a single real number,
\begin{align}
    \mathcal{M}^{(n)}_{\omega_r}=
      \frac{1}{\sqrt{2\pi\sigma^2}}
        \sum_{i}^M
        \sum_{j}^N (\mathbf{X}^n)_{ij}\,
        \mathrm{e}^{
          -\left(\omega_r-\omega_{ij}\right)^2/2\sigma^2}\,.
    \label{eq:resonance-analysis-measure}
\end{align}
In \cref{eq:resonance-analysis-measure}, we have introduced the coupling matrix $\mathbf{X}$, with
  \begin{equation}
    \mathbf{X}_{ij} =
    \begin{cases}
      1 & \text{if } \big(\op{H}_{Q-C}\big)_{ij} \neq 0\\
      0 & \text{otherwise}
    \end{cases}\,,
    \label{eq:coupling-matrix-elements}
  \end{equation}
to capture the transitions which are allowed by $\op{H}_{Q-C}$ in \cref{eq:tunable-coupler-hamiltonian}. 
$\mathcal{M}^{(n)}_{\omega_r}$ expresses the proximity of all relevant transition frequencies to a reference frequency $\omega_r$. The transition frequencies are given by $\omega_{ij}=|\tilde{E}_i-\tilde{E}_j|$ where $\tilde{E}_{i/j}$ denote eigenenergies 
of the undriven system. The transition frequencies change when varying the spectator frequency $\omega_3$. In order to determine the undriven eigenstates, we need to fix the value of the coupler frequency. 
We take a value close to the average drive and adjust it in order to match the spectrum perfectly. The index $j$ runs over all $N$ states of the truncated Hilbert space whereas the index $i$ is limited to the set of $M<N$ states consisting of all logical basis states of the three qubits plus those states that are purposefully involved in the gate protocol. To adjust the sensitivity of $\mathcal M^{(n)}_{\omega_r}$, we use a narrow Gaussian with width $\sigma$ and found $\frac{\sigma}{2\pi}=4\,\mathrm{MHz}$ to be a good choice to identify all relevant features.

We will start by analyzing  $\mathcal{M}^{(n=1)}_{\omega_r}$ for 
first order $X$-transitions. These require
the transition frequency to be equal to a frequency component of the drive, i.e., 
$\omega_\phi$ and $2\omega_\phi$ 
according to \cref{eq:time-dependent-freq}.
When treating the drive by perturbation theory, first order $X$-transitions occur not only in first order perturbation theory, as one would expect, but also in second order perturbation theory, where the combinations $XZ$ and $ZX$ contribute, as we detail in \cref{app:perturbation-theory-second-order}.
These lead to multi-photon processes, 
and we need to consider transitions with integer multiples of $\omega_\phi$ up to 4, see
\cref{app:higher-order-harmonics-excitation-in-coupled-transmon-systems} for a detailed discussion of the contributions at first and second order perturbation theory.

When going to second order $X$-transitions,  only contributions starting from second order perturbation theory are relevant. In addition to considering integer multiples of $\omega_\phi$ up to 4, as just explained above,  we anticipate that the diagonal part of the drive also gives rise to contributions from the next higher order perturbative term. Thus, we expect third order perturbation theory to  contribute second order $XX$-transitions which will result in combinations of three times the original drive frequencies $\omega_\phi$ and $2\omega_\phi$. Hence, we will consider integer multiples of $\omega_\phi$ up to 6 for $\mathcal{M}^{(n=2)}_{\omega_r}$.

\begin{figure*}
    \includegraphics[width=\textwidth]{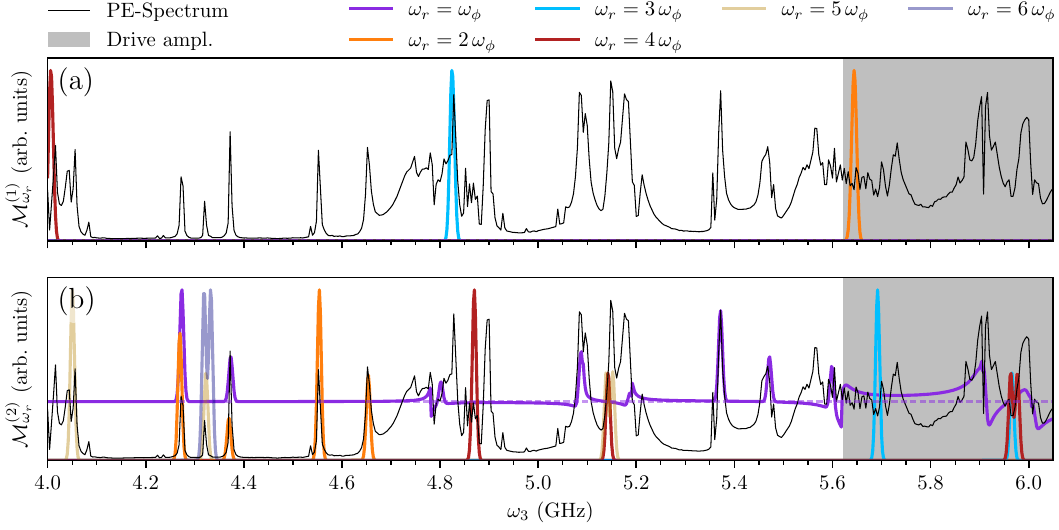}
    \caption{
    Comparison of the PE spectrum (black line) to the resonance measure $\mathcal{M}^{(n)}_{\omega_r}$,
      \cref{eq:resonance-analysis-measure}, for the CZ gate with
      $\alpha_3=100\,\mathrm{MHz}$, shown for various reference frequencies $\omega_r$ and $n=1$ (a) and $n=2$ (b).
      The dashed purple line indicates the offset of $\mathcal{M}^{(2)}_{\omega_r=\omega_\phi}$ originating from the transition used to implement the CZ gate.
    }
    \label{fig:resonance_analysis_ganzhorn}
\end{figure*}
\figureref{fig:resonance_analysis_ganzhorn} compares the resonance measure  $\mathcal{M}^{(n)}_{\omega_r}$ with the PE spectrum for first (a) and second order (b) $X$-transitions using
reference frequencies $\omega_r$ that are mulitples of $\omega_\phi$, as discussed above. In the calculation of the $\omega_{ij}$ in \cref{eq:resonance-analysis-measure},
$\omega_c$ was adjusted to $7.266\,\mathrm{GHz}$ (compared to an average of $\approx 7.25\,\mathrm{GHz}$) in order to improve alignment of $\mathcal{M}^{(n)}_{\omega_r}$ with the peaks in the PE spectrum.
All curves exhibit multiple peaks, which align well with peaks in the PE spectrum, indicating that drive-induced resonances represent a major source of crosstalk.
\figureref[(a)]{fig:resonance_analysis_ganzhorn}
displays only a single peak in $\mathcal{M}^{(1)}_{\omega_r}$ for a given reference frequency $\omega_r$. These correspond to drive-induced transitions between the tunable coupler and the spectator qubit. Such first order couplings are in general strong and population swaps between the tunable coupler and the spectator qubit occur even at large detunings of the spectator frequency from the resonance. This is reflected by the comparatively broad regions in the PE spectrum that indicate crosstalk around the first order resonances. In these regions, the gate protocol is severely impaired, which leads to a significantly reduced population transfer to $\ket{02k}$. This also explains the missing peak for the static resonance peak around $\omega_3=5.6\,\mathrm{GHz}$, discussed at the end of \cref{ssec:static-resonances}, 
which required contributions of $\ket{02k}$. 
In contrast, the peak for $\omega_r=4\omega_\phi$ in \figref[(a)]{fig:resonance_analysis_ganzhorn} is not very broad, since the resonance consists of two excitations with frequency $2\omega_\phi$. This frequency component in the drive $\omega_c(t)$ is already comparatively small.

Second order transitions account for the remaining peaks in the PE spectrum, cf. \figref[(b)]{fig:resonance_analysis_ganzhorn}.
In the CZ gate protocol, the drive frequency $\omega_\phi$ is taken to be (near-) resonant with the second-order transition $\ket{11k}\leftrightarrow\ket{02k}$. The corresponding resonance measure (purple line) exhibits a constant non-zero offset (dashed purple line). We estimate the offset from regions with zero crosstalk and plot it as dashed purple line.
Deviations from the constant offset indicate either additional transitions with  a frequency of $\omega_\phi$ or changes in the transition frequency of $\ket{11k}\leftrightarrow\ket{02k}$ due to dressing by the third qubit.
Both lead to unwanted crosstalk, as seen in the coincidence with large values of the PE spectrum.
In case of an additional transition, the deviations exhibit sharp peaks, for example around $\omega_3=4.25\,\mathrm{GHz}$ or $\omega_3=5.4\,\mathrm{GHz}$. In contrast, small deviations, especially negative ones, suggest a change of the  $\ket{11k}\leftrightarrow\ket{02k}$ transition frequency. This can lead to significant crosstalk, since the gate mechanism for the CZ-gate requires a precise phase accumulation when cycling to $\ket{02k}$ and back. It is confirmed in the PE spectrum, for example around $\omega_3=5.0\,\mathrm{GHz}$ or $\omega_3=5.8\,\mathrm{GHz}$ where
the small deviations in $\mathcal{M}_{\omega_r}$ correlate with large values of the PE spectrum.
Combining all observations, all features of the PE spectrum are rationalized in terms of the mechanisms that lead to crosstalk for different parameters.

\section{Analyzing-Crosstalk Mechanisms: Application to a \texorpdfstring{\sqrtiswap{}}{sqrt-iSWAP} Gate}
  \label{sec:analysing-crosstalk-mechanisms-iswap}

\begin{table}[tbp]
    \caption{Parameters for \sqrtiswap{} gate~\cite{McKayPRA16}.}
    \label{tab:parameters-table-mckay}
    \begin{ruledtabular}
      \begin{tabular}{llcr}
        \multirow{3}{4em}{Qubit 1}
          & Frequency & $\omega_1/2\pi$ & $5.8899\,\mathrm{GHz}$\\
          & Anharmonicity & $\alpha_1/2\pi$ & $324\,\mathrm{MHz}$\\
          & Coupling & $g_1/2\pi$ & $100.0\,\mathrm{MHz}$\\
      \colrule
        \multirow{3}{4em}{Qubit 2}
          & Frequency & $\omega_2/2\pi$ & $5.0311\,\mathrm{GHz}$\\
          & Anharmonicity & $\alpha_2/2\pi$ & $235\,\mathrm{MHz}$\\
          & Coupling & $g_2/2\pi$ & $71.4\,\mathrm{MHz}$\\
      \colrule
        \multirow{3}{4em}{Qubit 3}
          & Frequency & $\omega_3/2\pi$ & varied\\
          & Anharmonicity & $\alpha_3/2\pi$ & $100\,\mathrm{MHz}$\\
          & Coupling & $g_3/2\pi$ & $85.0\,\mathrm{MHz}$\\
      \colrule
        \multirow{2}{4em}{Tunable coupler}
          & Frequency & $\omega^{\mathrm{max}}_c/2\pi$ & $7.445\,\mathrm{GHz}$\\
          & Anharmonicity & $\alpha_c/2\pi$ & $230\,\mathrm{MHz}$\\
      \colrule
        \multirow{4}{4em}{Drive}
          & Offset & $\Theta$ & $-0.108$\\
          & Amplitude & $\delta$ & $0.155$\\
          & Frequency & $\omega_\phi/2\pi$ & $850.6\,\mathrm{MHz}$\\
          & On/Off flank width & $\sigma_t$ & $8.3\,\mathrm{ns}$\\
      \end{tabular}
    \end{ruledtabular}
  \end{table}
  
  \begin{figure*}
    \includegraphics[width=\textwidth]{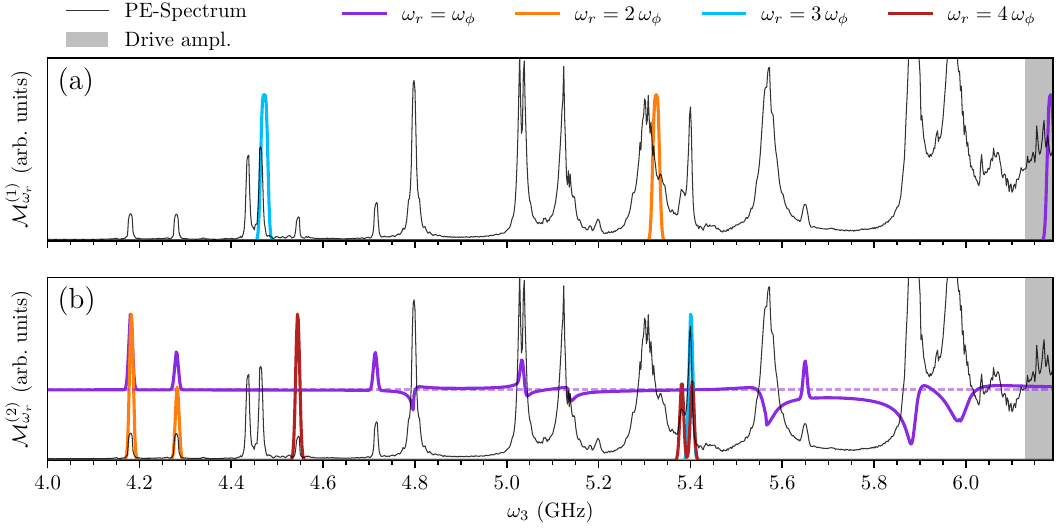}
    \caption{
      Perfect entangler spectrum (black line) and resonance analysis, quantified
      by $\mathcal{M}^{(n)}_{\omega_r}$, \cref{eq:resonance-analysis-measure} for
      a \sqrtiswap{} gate with $\alpha_3=100\,\mathrm{MHz}$ with $n=1$ (a) and $n=2$ (b).
      The dashed purple line indicates the resonance of $\omega_\phi$ in the
      system employed for implementing the \sqrtiswap{} gate.
    }
    \label{fig:resonance_analysis_mckay}
  \end{figure*}
  
As a second example, we analyze a \sqrtiswap{} gate, using the parameters from  Ref.~\cite{McKayPRA16}, given in \cref{tab:parameters-table-mckay}. Since many aspects are similar to the CZ-protocol, we focus on the main results. \figureref{fig:resonance_analysis_mckay} shows the spectra for the  \sqrtiswap{} gate analogously to \figref{fig:resonance_analysis_ganzhorn}. Again, the PE  spectrum exhibits several peaks, corresponding to the presence of unwanted crosstalk in the system. Overall, the spectrum is less densely peaked, compared with the CZ gate.
In addition to the PE spectrum, we also plot the resonance  measure $\mathcal{M}^{(n=1,2)}_{\omega_r}$, \cref{eq:resonance-analysis-measure}, for several reference frequencies. $\mathcal{M}_{\omega_r}$ is calculated with $\omega_c=7.000\,\mathrm{GHz}$, which we determined to yield the best match with the peaks of the PE spectrum. We show resonances with integer multiples of the flux frequency $\omega_\phi$ up to $4$. Peaks for larger multiples of $\omega_\phi$, as observed for the CZ gate, do not appear in the spectrum, and thus are omitted.  This is expected, due the smaller drive amplitude and shorter interaction time of the \sqrtiswap{}, which both attenuate the influence of higher order terms. For all reference frequencies, the curves for $\mathcal{M}_{\omega_r}$ exhibit several peaks, which align well with the peaks of the PE spectrum.

The \sqrtiswap{} gate is mediated by driving the second order transition $\ket{01}\leftrightarrow\ket{10}$ with a near-resonant drive. Similarly to the case of the CZ gate, this resonance is reflected as offset in $\mathcal{M}^{(2)}_{\omega_r=\omega_\phi}$  (purple line in \figref[(b)]{fig:resonance_analysis_mckay}). We expect the \sqrtiswap{} protocol to be more robust to small changes of the transition frequency, compared to the CZ gate, since it just implements
a population swap between $\ket{01k}$ and $\ket{10k}$ and does not require the precise accumulation of a relative phase. This robustness is indeed seen in the PE spectrum, \eg around $4.9\,\mathrm{GHz}$ in \figref[(b)]{fig:resonance_analysis_mckay}. Here, a moderate deviation from the expected offset (dashed line) leads to only a comparatively small increase of the PE spectrum.

Finally, we address the only peak in the PE spectrum that does not seem to have any correspondence with a peak in $\mathcal{M}^{(n=1,2)}_{\omega_r}$ --- the  
left of the two peaks close to $4.45\,\mathrm{GHz}$. The right peak coincides with a resonance due to $3\omega_\phi$ (light blue curve). Upon closer investigation, both peaks are caused by the same crosstalk mechanism, namely a swap of excitation between the spectator qubit and the tunable coupler. Their peak separation originates from a shift of the spectator eigenenergies, that depends on the state of the other two qubits. However, the splitting is much more pronounced than expected. This can be attributed to the drive leading to a time-dependent AC-Stark shift of the associated levels. Time-dependent effects are not captured when determining the eigenergies needed to calculate $\mathcal{M}_{\omega_r}$ in \cref{eq:resonance-analysis-measure}.
We have confirmed this interpretation by comparing spectra for various drive strengths (data not shown), which indeed shows a correlation of the peak splitting with the drive amplitude.


\section{Experimental Implementation}
  \label{sec:experimental-implementation}

Finally, we suggest a protocol to experimentally measure the PE spectrum, in order to identify the various crosstalk mechanisms discussed above on a real device.
To determine the values of the combined functional $J$, \cref{eq:full-functional}, in an experiment, a naive approach would be to determine the full three-qubit gate directly, for instance, through a process tomography for three qubits.
A more efficient protocol allows for reconstructing the gate by only evolving and measuring $N+1$ states~\cite{Romer25}, where $N$ is the gate dimension, \ie $N=8$ for 3-qubit gates here. 
Each of these states is prepared individually, the gate is applied to it, and the resulting states are determined by state tomography.
Note, that it is sufficient to perform a partial state tomography in the logical
subspace, since the desired gate is restricted to this subspace.
From the outcome of the state tomographies, we can reconstruct the quantum gate~\cite{Romer25}.
In other words, it is only necessary to apply  the gate to all states of a logic basis plus one additional state in a rotated (mutually unbiased) basis~\cite{Romer25}.

Since the two-qubit gate should not affect the state of the spectator qubit,
the targeted three-qubit gate can be written as block diagonal matrix,
\begin{align}
    U_{3q} = U_{2q}\otimes \openone =
    \begin{pmatrix}
        U_{2q} & 0\\
        0 & U_{2q}
    \end{pmatrix}\,,
    \label{eq:block-diag-matrix}
\end{align}
where $U_{2q}$ is a two-qubit gate between the first two qubits.
This property allows us to reduce the effort to determining two two-qubit
gates: One for the spectator qubit being in $\ket{0}$ and one for the
spectator qubit being in $\ket{1}$.
This in turn has the advantage, that the effort for the state tomographies can be further reduced to a two-qubit subspace, i.e. $N=4$, instead of $N=8$.
After determining both two-qubit gates, $J$ can be calculated according to
\cref{eq:full-functional}.
Finally, the weights $w_S$ and $w_U$ in \cref{eq:full-functional} and
\cref{eq:perfect-entangler-functional}, respectively, can be adjusted to
maximize the contrast of the spectrum.

Importantly, while the relative phase between the two-qubit
gates does not change the perfect entangler spectrum, a non-vanishing relative
phase will induce an undesired local $Z$-rotation for the third qubit. In order to obtain a crosstalk-free gate, it will be necessary to determine this relative phase and compensate it by a local $Z$-gate on the
spectator. This can be done by preparing an initial product state, with the spectator qubit
in $(\ket{0}+\ket{1})/\sqrt{2}$.
Since the two two-qubit gates are the same
(up to a relative phase), application of the gate protocol leaves the three-qubit state separable with respect to the spectator qubit but rotates the spectator qubit, with 
$\varphi$ the relative phase between the two-qubit gates.
Thus, measuring the $\op{\sigma}_x$ and $\op{\sigma}_y$ components of the spectator qubit state yields this phase.

Although the protocol described above allows for the measurement of the
perfect entangler functional, it does not yield the minimal value of the
perfect entangler functional over time.
Instead, in its simplest form, the protocol only allows to determine the
functional value at a fixed final time $T$.
In order to estimate the impact of this limitation, we compare the two cases
for the CZ gate.
To this end, we determine the value of $J$ at a fixed time of
$T=690\,\mathrm{ns}$, which provided the largest contrast for the spectrum, 
\begin{figure}
  \includegraphics[width=\columnwidth]{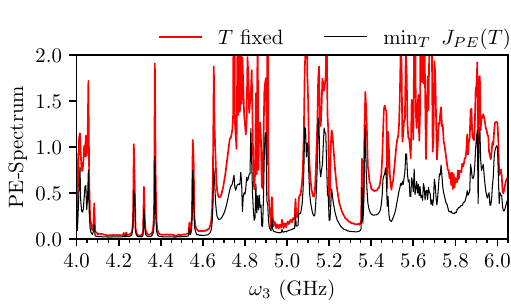}
  \caption{
    Comparison of spectra, obtained with the fixed-time and the
    minimization-over-time method.
    The black curve shows the perfect entangler spectrum of the CZ
    gate from \cref{fig:pe_spectrum}.
    The red curve shows a similar spectrum without minimization over time,
    with $T=690\,\mathrm{ns}$.
  }
  \label{fig:comparison_fixed_T_vs_min_JT}
\end{figure}
cf. the red curve in  \figref{fig:comparison_fixed_T_vs_min_JT}.
When comparing this to using the minimal value of $J$ over time
(black curve in \figref{fig:comparison_fixed_T_vs_min_JT}), the fixed-time spectrum shows much larger peaks, and some noise is introduced, for example oscillations around $5.0\,\mathrm{GHz}$.
However, all main features of the spectrum are preserved.
This suggests that the experimentally much more feasible measurement of the
spectrum at a final time $T$ is sufficient for detecting the relevant crosstalk
mechanisms.


\section{Conclusions \& Outlook}
  \label{sec:conclustion-and-outlook}

We have introduced the perfect entangler spectrum, a versatile tool for
analyzing crosstalk, and  shown how to use it for a parametrically driven tunable coupler architecture, simultaneously coupling more than two qubits. The perfect entangler spectrum 
allows for precise quantification of gate crosstalk and the influence of the different states of spectator qubits. In more detail, 
it identifies qubit frequencies which do not lead to crosstalk, facilitating the design of tunable coupler architectures.

Moreover, we have shown how to leverage the perfect entanglers spectrum to identify the crosstalk mechanisms. In the examples we have considered, these can be divided into static and drive-modulated resonances.
The former consist of unwanted resonances between transition frequencies which can already be inferred from the bare system Hamiltonian and do not necessitate a dedicated spectroscopy.
The drive modulated resonances, on the other hand, originate from unwanted
transitions induced by the drive.
In particular, we find that multiphoton processes lead to unexpected
crosstalk for certain parameter regimes.

Finally, we have outlined a protocol to
experimentally measure the perfect entangler spectrum, showing that it is not just a useful tool for theoretical and computational analysis but also practically feasible. It requires, for each spectator frequency, two two-qubit gate tomographies plus the determination of a single-qubit rotation phase.

Architectures coupling even more qubits with
a single tunable coupler have been suggested in the literature~\cite{Huber24}.
While we restrict our study to three coupled qubits, our insights on the
crosstalk mechanisms can be generalized to such more complex systems.
This can aid the selection of parameter regimes when adding more qubits
to a single coupler.
Our crosstalk spectroscopy would also be sensitive to more complex error channels~\cite{Kishmar25,Chowdhury25}.
Further, our approach is complementary to crosstalk tomography via random unitary circuits~\cite{HelsenNC23,ManginiPRR24} that diagnoses occurrence of crosstalk without the capability of locating the source or identifying the mechanism causing it.
Additionally, while we have considered entangling two-qubit operations in our examples, our approach to crosstalk spectroscopy is easily extended to single qubits as well. This is easily done by optimizing for a set of fixed single qubit gates, e.g. using the average fidelity instead of the perfect entangler functional. The approach can therefore be used to analyze crosstalk for a complete universal set of gates.
As a next step it will be interesting to use the perfect entangler spectrum as a diagnostic tool when varying the gate protocol. For example, optimal control methods~\cite{KochEPJQT22} can be used to mitigate crosstalk dynamically when frequency crowding prevents the simple mitigation strategy of choosing the right qubit parameters to be effective. The perfect entangler spectrum will provide a suitable figure of merit for such optimizations.


\begin{acknowledgments}
  Funding from the Deutsche Forschungsgemeinschaft (DFG, German Research
  Foundation)–Projektnummer 277101999–TRR 183 (project C05) and from the German
  Federal Ministry of Education and Research (BMBF) within the project QCStack
  (13N15929) is gratefully acknowledged.
\end{acknowledgments}

\appendix

\section{Local invariants of two-qubit gates}
\label{app:LI}

For completeness, we recall here the classification of two-qubit gates in terms of local equivalence classes~\cite{MakhlinQIP02,ZhangPRA03}. Any two-qubit gate is an element of $SU(4)$, and 
to assess whether a two-qubit gate is a ``perfect entangler'', one can make use of the local equivalence classes of $SU(4)$. They arise from the fact that any two-qubit gate can be written as
\begin{align}
    \op{U} = \op{k}_1
      \exp\bigg[ \frac{i}{2}\big(
        c_1\op{\sigma}_x\otimes\op{\sigma}_x +
        c_2\op{\sigma}_y\otimes\op{\sigma}_y +
        c_3\op{\sigma}_z\otimes\op{\sigma}_z
      \big)\bigg]
    \op{k}_2\,.
    \label{eq:cartan-decomposition}
\end{align}
Here $\op{k}_1$ and $\op{k}_2$ are local operations, i.e.,  $\op{k}_i\in\mathrm{SU}(2)\otimes\mathrm{SU}(2)$, and the non-local part of the gate is represented by the exponential. It is fully characterized by only three real numbers ($c_1$, $c_2$, $c_3$) -- the Weyl coordinates. Accounting for the symmetries of the exponential in \cref{eq:cartan-decomposition} reduces the cube, spanned by $c_{1,2,3}\in (0,\pi)$, to the so called Weyl chamber. Gates sharing the same coordinates in the Weyl chamber belong to the same local equivalence class and can be converted into one another with local operations. The best known examples of such local equivalence are CNOT and the controlled phasegate (CZ), their Weyl coordinates are $c_1=\frac{\pi}{2}$, $c_2=0$, $c_3=0$.

However, determining the Weyl coordinates of a gate can be difficult, and it is more practical to classify gates by their local invariants~\cite{MakhlinQIP02,ZhangPRA03}
  \begin{align}
      g_1 &= \frac{1}{16}\Real\Big[\!\Tr^2({M})\Big],
      \qquad
      g_2 = \frac{1}{16}\Imag\Big[\!\Tr^2({M})\Big],
    \notag\\
      g_3 &= \frac{1}{4}\Big[\!\Tr^2({M})-\Tr\big({M^2}\big)\Big]\,,
  \end{align}
where $M = U_B^TU_B$ and $U_B$ corresponds to the two-qubit gate in the  Bell basis. 
  
Inside the Weyl chamber, all perfectly entangling gates form a polyhedron, also  called perfect entangler polyhedron. Given this geometric picture, a given gate is either inside the polyhedron or has a well-defined distance from it~\cite{WattsPRA15}. This distance can be expressed in terms of the Weyl coordinates $c_1$, $c_2$, $c_3$ or the local invariants $g_1$, $g_2$, $g_3$ where use of the latter in calculating the distance comes with the advantage of smoothness of the function. 

The local invariants can be used to design protocols that yield a representative of a local equivalence class rather than a specific gate~\cite{MuellerPRA11} or an arbitrary perfect entangler~\cite{WattsPRA15,GoerzPRA15,GoerzPRA15}. They can also be used in parametrized gate design to improve upon existing drive sequences~\cite{Sugawara25}.

\section{Drive in the dressed basis}
  \label{app:diagonal-drive-transitions}

In order to understand the effect of the drive in \cref{eq:tunable-coupler-hamiltonian}, we switch to the dressed basis, where  the dressing is due to the interaction between qubits and coupler with strength $g$.
Also, it is useful to split the Hamiltonian \eqref{eq:tunable-coupler-hamiltonian} into time-independent and time-dependent parts, 
\begin{align}
 \op{H}(t) &= \op{H}_0 + u(t)\op{H}_1\,.
 \label{eq:abstract-drive-Hamiltonian}
\end{align}
We follow~\cite{McKayPRA16} and treat the $\op{H}_0$ as the Hamiltonian from \cref{eq:tunable-coupler-hamiltonian} with $\delta=0$ in~\cref{eq:time-dependent-pulse}.
With this, the time-dependent part is given as 
$\op{H}_1=\omega_c^\mathrm{max}\op{b}^\dagger\op{b}$
with $u(t)=\sqrt{\big|\cos{\big(\pi\,\Phi(t)\big)}\big|} - \sqrt{\big|\cos{\big(\pi\,\Theta\big)}\big|}$.

For simplicity, we consider the subsystem of only one fixed-frequency transmon and the coupler and treat both as two-level systems. 
The drift and control Hamiltonians in \cref{eq:abstract-drive-Hamiltonian} are then given by
\begin{subequations}
\begin{align}
  \label{eq:simplified-coupling-model}
  \op{H}_0 &= \frac{\omega_a}{2} \op{\sigma}_z \otimes \op{\openone}
  + \frac{\omega_b}{2} \op{\openone} \otimes \op{\sigma}_z
  + g(\op{\sigma}_+ \otimes \op{\sigma}_-
  + \op{\sigma}_- \otimes \op{\sigma}_+) \notag\\
  &= \begin{pmatrix}
        \frac{\omega_b+\omega_a}{2} & 0 & 0 & 0 \\
        0 & \frac{\Delta\omega}{2} & g & 0 \\
        0 & g & -\frac{\Delta\omega}{2} & 0 \\
        0 & 0 & 0 &  -\frac{\omega_a+\omega_b}{2}
    \end{pmatrix}\,,\\
    \op{H}_1 &= \frac{1}{2}\op{\sigma}_z \otimes \op{\openone} \,,
    \label{eq:H1}
\end{align}
\end{subequations}
where $\Delta\omega = \omega_a-\omega_b$. Note that we employ the rotating wave approximation, neglecting the terms $\op{\sigma}_+\otimes\op{\sigma}_+$ and $\op{\sigma}_-\otimes\op{\sigma}_-$, which is justified for the parameter values used in the main text.
The drift Hamiltonian $\op{H}_0$ is block-diagonal and the coupling only occurs in the central subspace spanned by $|01\rangle$ and $|10\rangle$.
Focusing on this subspace, we can rewrite the Hamiltonian in the subspace as
\begin{subequations}  
  \begin{align}
  \label{eq:qubit_subspace_hamiltonian_h0}
  \op{\tilde{H}}_0 &= \Delta\omega \op{\tilde{\sigma}}_z + g\op{\tilde{\sigma}}_x
    = \begin{pmatrix}
        \frac{\Delta\omega}{2} & g \\
        g & -\frac{\Delta\omega}{2} \\
      \end{pmatrix}
\end{align}
with  eigenenergies $E_\pm = \pm
\sqrt{\left(\frac{\Delta\omega}{2}\right)^2 + g^2} \equiv \pm\Omega$ and
\begin{align}
    \label{eq:qubit_subspace_hamiltonian_h1}
    \op{\tilde{H}}_1 &=
    \frac{1}{2}\op{\tilde{\sigma}}_z\,.
\end{align}
\end{subequations}
In the eigenbasis of $\op{\tilde{H}}_0$, the control term becomes
\begin{align}
  \op{\tilde{H}}^\mathrm{dressed}_1 =
    \frac{\Delta\omega}{2\Omega}\op{\tilde{\sigma}}_z
    + \frac{g}{\Omega}\op{\tilde{\sigma}}_x
  \label{eq:h1-dressed}
\end{align}
i.e., effectively the drive has both a diagonal and an off-diagonal component.
When translating this result back to the total space of the qubit-coupler system, we find the control Hamiltonian~\eqref{eq:H1} to be separated into a diagonal and an off-diagonal component,
\begin{align}
    \op{H}^\mathrm{dressed}_1 = \begin{pmatrix}
        \frac{1}{2} & 0 & 0 & 0 \\
        0 & \frac{\Delta\omega}{2\Omega} & \frac{g}{\Omega} & 0 \\
        0 & \frac{g}{\Omega} & -\frac{\Delta\omega}{2\Omega} & 0 \\
        0 & 0 & 0 & -\frac{1}{2}
    \end{pmatrix} \equiv \op{Z} + \op{X}\,,
    \label{eq:h1-dressed-full}
\end{align}
with $\op{X}=g/\Omega\big(\op{\sigma}_+ \otimes \op{\sigma}_- + \op{\sigma}_- \otimes \op{\sigma}_+\big)$ representing the off-diagonal contribution. It vanishes for weak coupling.

\section{Second order perturbation theory in the drive}
  \label{app:perturbation-theory-second-order}
We provide a brief analysis of the transitions induced by a modulation of the coupler frequency $\omega_c(t)$ in \cref{eq:tunable-coupler-hamiltonian}.
In order to understand the effect of the drive, we switch to a basis which is dressed with respect to the interaction between qubits and coupler, $\op{H}_{Q-C}$ in \cref{eq:tunable-coupler-hamiltonian}. While this is challenging in general, we know from \cref{app:diagonal-drive-transitions} that the effective drive  Hamiltonian $\op{H}_1$ comprises a diagonal and an off-diagonal part in the dressed basis.
Similarly to \eqref{eq:h1-dressed-full}, we refer to the diagonal term as $\op{Z}$ and the off-diagonal term as $\op{X}$,
\begin{align}
    \op{H}_1 = \op{Z} + \op{X}\,,
    \label{eq:H_1_simplified}
\end{align}
with $\op{X}_{ij} \propto \Big(\sum_k  \big(\op{b}\op{a}_k^\dagger+\op{b}^\dagger\op{a}_k\big)\Big)_{ij}$.
As in \cref{eq:h1-dressed-full}, the matrix elements of $\op{X}$ are small compared to those of $\op{Z}$ if the couplings $g_i$ are small compared to the detunings between qubit and coupler, $g_i\ll\Delta\omega_i$. In our examples, they are smaller by about a factor of 10. 

To examine which transitions are induced by the drive in \cref{eq:H_1_simplified}, we employ time-dependent perturbation theory up to second order. The transition amplitude $P_{n\rightarrow m}$ between two levels $\ket{n}$ and $\ket{m}$ is then  given by
\begin{widetext}
  \begin{align}
  \label{eq:second_order_perturbation_theory_toy_model}
   P^{(2)}_{n\rightarrow m} &= \bigg|\delta_{mn} - iV_{mn}\int_0^t \mathrm{d}t'\ 
     \e{i\omega_{mn}t'} u(t')
     -  \sum_k V_{mk}V_{kn}\int_0^t \mathrm{d}t' \int_0^{t'} \mathrm{d}t''\ 
     \e{i\omega_{mk}t'}\e{i\omega_{kn}t''}  u(t')u(t'')
     \bigg|\,.
 \end{align}
\end{widetext}
Here, $V_{ij} = \braket{i|\op{H}_1|j}$ and $\omega_{ij} = {E}_i - {E}_j$, where ${E}_i$ are the eigenenergies and $\ket{i}$, $\ket{j}$ the eigenstates, dressed with respect to the qubit-coupler interaction $\op{H}_{Q-C}$.
Since we consider transitions between distinct levels, $\ket{m}\neq \ket{n}$, the first term on the rhs.\ of \cref{eq:second_order_perturbation_theory_toy_model} does not contribute.
The second term, describing first order perturbations, is non-zero only if $V_{mn}\neq 0$.
With the definition of $\op{H}_1$, \cf \cref{eq:H_1_simplified}, we find
\begin{align}
  \label{eq:transition-amplitude-definition}
  V_{mn} = \braket{m|\op{H}_1|n} = \underbrace{\braket{m|\op{Z}|n}}_{=0\text{ for }m\neq n}+\braket{m|\op{X}|n}\,,
\end{align}
\ie within first order perturbation theory, transitions between two states $\ket{m}$ and $\ket{n}$ occur only of they are connected by $\op{X}$.
The last term in \cref{eq:second_order_perturbation_theory_toy_model}, obtained in second order perturbation theory, results in transitions if $\sum_k V_{mk}V_{kn}$ is non-zero.
The sum over all possible states $\ket{k}$ enables transitions between states which are coupled by $\op{X}\op{X}$. 
The second order term includes also transitions between states which are coupled by $\op{X}$, \ie first order $X$-couplings, since the drive Hamiltonian contains the diagonal term $\op{Z}$. 
In other words, these correspond to transitions between states coupled by $\op{X}\op{Z}$ or  $\op{Z}\op{X}$.
Recall that the matrix elements of $X$ are small compared to those of $Z$. Therefore transitions due to $ZX$ or $XZ$ dominate over those due to $XX$.
  
\section{Higher order harmonic excitation in coupled transmon systems}
  \label{app:higher-order-harmonics-excitation-in-coupled-transmon-systems}

As illustrated in \figref[(a)]{fig:resonance_analysis_ganzhorn} and \figref[(a)]{fig:resonance_analysis_mckay}, we observe several crosstalk contributions, corresponding to first order $X$-transitions between the tunable coupler and the third qubit.
For instance, the peak in \figref[(a)]{fig:resonance_analysis_mckay} around $\omega_3 \approx 4.46\,\mathrm{GHz}$ corresponds to a 
swap between the spectator qubit and the tunable coupler, but with a transition frequency of $\omega_r=3\omega_\phi$.
This swap occurs, despite the drive not containing the frequency component $3\omega_\phi$, but only $\omega_\phi$ and $2\omega_\phi$.
In the following, we will exemplify how such a transition is possible, due to second order $XZ$ interactions between the third qubit and the coupler.

To understand the swap, we use the simplified model of \cref{app:diagonal-drive-transitions}
and start with the effective drive in \cref{eq:h1-dressed}. We calculate $P^{(2)}_{n\rightarrow m}$ with $n\neq m$ and
$n,m\in\{\tilde{0},\tilde{1}\}$.
The first order contribution in
\cref{eq:second_order_perturbation_theory_toy_model} yields a transition
proportional to the Fourier transform of $u(t)$: $\mathcal{F}\big[u\big](\omega) = \int_0^t \mathrm{d}t'\ \e{i\omega_{mn}t'} u(t')$.
In \cref{eq:h1-dressed},  $V_{mn} = g/\Omega$ for $m\neq n$, and thus the first order perturbation term becomes $i(g/\Omega)\mathcal{F}\big[u\big](\omega_{mn})$.
Assuming a pulse shape $u(t) \approx a+b\cos(\omega t) + c\cos(2\omega t)$, the first order term is non-zero only if $\omega_{mn} \approx \omega$ or $\omega_{mn} \approx 2\omega$.
This does not yet explain the transition between levels with a transition frequency of $\omega_{mn} = 3\omega$ seen in the perfect entangler spectrum, \ie it is necessary to include second order contributions.

The second order contribution is more complicated, due to the
time ordering in \cref{eq:second_order_perturbation_theory_toy_model}.
As discussed in \cref{app:perturbation-theory-second-order},  transitions due to $ZX$ and $XZ$ dominate the second order contribution.
Expanding the sum in \cref{eq:second_order_perturbation_theory_toy_model} yields two conditions which we
treat independently.
The first, corresponding to $ZX$ or $n=k$, yields
\begin{align}\nonumber
  \int_0^t & \mathrm{d}t' \int_0^{t'} \mathrm{d}t''
    \e{i\omega_{mn}t'} V_{mn}V_{nn}u(t')u(t'')
  \\\nonumber
    &=
    g\frac{\Delta\omega}{2\Omega^2}\int_0^t \mathrm{d}t'
      \e{i\omega_{mn}t'} u(t')\int_0^{t'} \mathrm{d}t''
     (-1)^n u(t'')
  \\
    &=
    (-1)^n g\frac{\Delta\omega}{2\Omega^2}\int_0^t \mathrm{d}t'
     \e{i\omega_{mn}t'} u(t') W(t')
  \,,\label{eq:ZX-contrib}
\end{align}
where we have utilized $\omega_{nn}=0$ and $V_{nn}=(-1)^n\frac{\Delta\omega}{2\Omega}$ from \cref{eq:h1-dressed}. Further, we have used the
antiderivative $W(s) = \int_0^{s}\mathrm{d}s' u(s')$.
Similarly, for $XZ$ resp. $m=k$,
\begin{align}\nonumber
  \int_0^t & \mathrm{d}t'' \int_{t''}^{t} \mathrm{d}t'
    \e{i\omega_{mn}t''} V_{mm}V_{mn}u(t')u(t'')
  \\ \nonumber
    &=
    g\frac{\Delta\omega}{2\Omega^2}\int_0^t \mathrm{d}t'' \e{i\omega_{mn}t''} u(t'')
      \int_{t''}^{t} \mathrm{d}t' (-1)^m u(t')
  \\\nonumber
    &=
    (-1)^m g\frac{\Delta\omega}{2\Omega^2}
      \int_0^t \mathrm{d}t'' \e{i\omega_{mn}t''} u(t'')
      \big[W(t')\big]^{t}_{t''}
  \\
    &=
    -(-1)^m g\frac{\Delta\omega}{2\Omega^2}
      \int_0^t \mathrm{d}t'' \e{i\omega_{mn}t''} u(t'')
      W(t'')
  \notag\\
    &\qquad +
    (-1)^m g\frac{\Delta\omega}{2\Omega^2}
      W(t)\int_0^t \mathrm{d}t'' \e{i\omega_{mn}t''} u(t'')
  \,,\label{eq:XZ-contrib}
\end{align}
where we have used the identity $\int_0^T\mathrm{d}t'\int_0^{t'}\mathrm{d}t''
f(t')g(t'') = \int_0^T\mathrm{d}t''\int_{t''}^{T}\mathrm{d}t' f(t')g(t'')$.
Combining Eqs.~\eqref{eq:ZX-contrib} and~\eqref{eq:XZ-contrib}, 
and letting $t\rightarrow\infty$, the dominant contributions in second order are obtained as 
\begin{align}\nonumber
  &\int_0^\infty \mathrm{d}t' \int_0^{t'} \mathrm{d}t'' \sum_k
  \e{i\omega_{mk}t'}\e{i\omega_{kn}t''} V_{mk}V_{kn}u(t')u(t'')
  \\\nonumber
  &=\big((-1)^n-(-1)^m\big)g\frac{\Delta\omega}{2\Omega^2}
      \int_0^\infty \mathrm{d}t'' \e{i\omega_{mn}t''} u(t'')
      W(t'')
  \notag\\\nonumber
  &\qquad +
    (-1)^m g\frac{\Delta\omega}{2\Omega^2} \tilde{W}
      \int_0^\infty \mathrm{d}t'' \e{i\omega_{mn}t''} u(t'')
  \\
  &=\notag\\
  &\frac{(-1)^m g\Delta\omega}{2\Omega^2}\bigg\{
    \tilde{W} \mathcal{F}\big[u(t)\big](\omega_{mn})
    - \mathcal{F}\big[u(t)W(t)\big](\omega_{mn})
  \bigg\} \,,
\end{align}
where $\tilde W=\lim_{t\rightarrow \infty}W(t)$
and we have used  $ (-1)^n-(-1)^m = -(-1)^m$ for $m\neq n$.
They contain the Fourier transform of $u(t)W(t)$ at the
transition frequency $\omega_{mn}$.
Assuming the same pulse shape as above, $u(t) \approx a+b\cos(\omega t) + c\cos(2\omega t)$,
$u(t)W(t)$ has frequency components $\omega$, $2\omega$,
$3\omega$ and $4\omega$, which explains the higher order harmonics observed in the PE spectrum.
%
%


\bibliography{references}

\end{document}